# Enhanced deep-freezing magneto- and elasto-caloric effects by modifying lattice anharmonicity and electronic structures

Xiao-Ming Huang[1,2,3,4,9], Ying Zhao[4,9], Xiaowen Hao[1,2,3], Hua-You Xiang[4], Jin-Han Yang[4], Chin-Wei Wang[5], Wenyun Yang[6], Cuiping Zhang[1,2,7], Binru Zhao[7], Jie Ma[7], Zongbin Li[4], Yafei Kuang[8], Liang Zuo[4], Xin Tong[1,2,3,*], Hai-Le Yan[4,*], Qingyong Ren[1,2,3,*]

[1]Institute of High Energy Physics, Chinese Academy of Sciences, Beijing 100049, China.

[2]Spallation Neutron Source Science Center, Dongguan 523803, China.

[3]Guangdong Provincial Key Laboratory of Extreme Conditions, Dongguan, 523803, China.

[4]Key Laboratory for Anisotropy and Texture of Materials (Ministry of Education), School of Material Science and Engineering, Northeastern University, Shenyang 110819, China.

[5]National Synchrotron Radiation Research, Hsinchu 300092, Taiwan.

[6]State Key Laboratory for Artificial Microstructure and Mesoscopic Physics and School of Physics, Peking University, Beijing 100871, China.

[7] Key Laboratory of Artificial Structures and Quantum Control, School of Physics and Astronomy, Shanghai Jiao Tong University, Shanghai 200240, China.

[8]School of Materials Science & Engineering, Taiyuan University of Science and Technology, Taiyuan 030024, China.

[9]These authors contributed equally: Xiao-Ming Huang, Ying Zhao.

*Corresponding authors:

Email: tongxin@ihep.ac.cn; yanhaile@mail.neu.edu.cn; renqy@ihep.ac.cn

ABSTRACT

Designing the high-performance magneto- or elasto-caloric effect in NiMnIn-based alloys with spin-lattice coupling in a deep-freezing temperature range of ~200 K to ~255 K is challenging due to the limited lattice entropy change $|\Delta S_{lat}|$ and large negative contribution of magnetic entropy change $|\Delta S_{mag}|$ during phase transitions. In this work, we systematically study the first-order magneto-structural transition in NiMnIn-based alloys by *in-situ* microstructural characterizations, physical property measurements, and first-principles calculations. A multi-element alloying strategy involving Cu and Ga co-doping is proposed to manipulate the phase transition. The co-doping reduces the lattice anharmonicity and thermal expansion coefficient of the martensitic phase, leading to an increase in the unit cell volume change $\Delta V$ and $|\Delta S_{lat}|$. It also modifies the electronic density of states, causing a decrease in the magnetization change $\Delta M$ and $|\Delta S_{mag}|$. The relief of the lattice mismatch reduces hysteresis losses in the refrigeration cycle. These synergetic effects yield excellent magneto- and elasto-caloric effects, with the effective magnetocaloric refrigeration capacity reaching up to ~182 J kg$^{-1}$ under the magnetic field of 5 T or an adiabatic temperature change of −4 K under a low field of 1.5 T and the elastocaloric coefficient of performance to ~30 or an adiabatic temperature change of −7 K with the strain of 5% at ~230 K, offering a potential solution for solid-state deep-freezing refrigeration.

# 1. Introduction

Solid-state refrigeration technology has received extensive attention due to its high efficiency and environmental friendliness. Solid-state refrigeration systems can be achieved with one or more of the following caloric effects: magnetocaloric (MCE), elastocaloric (eCE), barocaloric (BCE), and electrocaloric (ECE) effects, corresponding to large entropy changes induced by the external stimuli of magnetic field, uniaxial stress, hydrostatic pressure, and electric field, respectively [1, 2]. Materials with strong spin-lattice coupling (SLC) have large magneto-structural responses to magnetic field and uniaxial stress simultaneously, making them a group of promising MCE, eCE, and/or multicaloric candidates. As summarized in Fig. 1, typical SLC materials for solid-state caloric effects include LaFeSi- [3], GdSiGe- [4], MnAs- [5], MnNiGe- [6, 7], FeRh- [8], NiCoMnTi- [9, 10], NiFeGa- [11-13], NiMnX (X = Ga, Sn, Sb, In)-based [14-43] alloys, and so on. At present, the research on these materials is mainly aimed at daily use scenarios near the freezing point temperature, such as from −18°C (255 K) to 26°C (299 K). In contrast, there is relatively little research on lower temperature ranges, although refrigeration in a deep-freezing temperature range, *e.g.*, from ~200 K to ~255 K, is necessary for many special applications, such as biomedical engineering, exploration of space, seismic monitoring, *etc* [14, 44].

Nonetheless, the Ni–Mn–Z (Z is the *p*-block elements of In, Sn, and Sb) Heusler alloys have an easily driving magnetic martensitic phase transition with a wide operating temperature window, enabling to achieve deep-freezing refrigeration. Taking the ternary NiMnIn-based compounds as an example, the martensitic transformation temperature, $T_M$, could be easily tuned from 400 K to a lower temperature of 160 K by adjusting the valence electron concentration [45]. The austenitic phase always has a ferromagnetic configuration below the Curie temperature, $T_C^A$ [45], while the martensitic phase has an antiferromagnetic or 'weak magnetic' configuration [46]. When $T_M$ is lower than $T_C^A$, strong spin-lattice coupling is established, leading to a magnetic field- or uniaxial stress-driven magneto-structural transformation from a low-$T$ weak magnetic martensitic phase to a high-$T$ ferromagnetic austenitic phase.

However, designing high-performance NiMnIn-based magneto- or elasto-caloric materials for deep-freezing applications remains a very challenging task. One of the main difficulties is to obtain a large total phase transition entropy change, $\Delta S_{tr}$, which determines the upper limits of the adiabatic temperature change, $\Delta T_{ad}$, and of the isothermal entropy change, $\Delta S_{iso}$. $\Delta S_{tr}$ for SLC materials mainly composes the contributions from lattice vibrations or phonons, $\Delta S_{lat}$, and spin vibrations or magnons, $\Delta S_{mag}$, while the electronic contribution, $\Delta S_{elec}$, is almost negligible. Specifically for NiMnIn-based alloys, $\Delta S_{lat}$ makes a main contribution while $\Delta S_{mag}$ plays a negative role [47]. Therefore, a large $\Delta S_{tr}$ requires a large $|\Delta S_{lat}|$ but a small $|\Delta S_{mag}|$. The magnitude of $|\Delta S_{lat}|$ depends on the unit cell volume change ratio, $\Delta V/V_0$, during the phase transition, while the change in magnetization, $\Delta M$, determines $|\Delta S_{mag}|$. For most NiMnIn-based alloys, $\Delta M$ increases with decreasing temperature because the magnetization for the austenite phase increases as the temperature moves away from $T_C^A$ [45]. This would increase $|\Delta S_{mag}|$ and diminish $\Delta S_{tr}$ when pursuing a lower working temperature. Although it has been reported that doping Cu at the Mn sites can reduce $\Delta M$ and $|\Delta S_{mag}|$, this is at the expense of raising $T_M$ [48]. On the other hand, alloying with Ga in NiMnIn-based compounds can expand $\Delta V$, thereby increasing $|\Delta S_{lat}|$ and net $\Delta S_{tr}$, but this is still at the expense of raising $T_M$ [28]. All these factors make it fairly difficult to enlarge $\Delta S_{tr}$ by increasing $|\Delta S_{lat}|$ and suppressing $|\Delta S_{mag}|$ to pursue high-performance magneto- and elasto-caloric effects in a lower temperature range.

In this work, we proposed a multi-element alloying strategy to realize a high-performance deep-freezing magneto- and elasto-caloric effect in the NiMnIn-based Heulser alloys. Ga and Cu were selected to partially replace In and Mn, respectively, while the content of In was carefully balanced. This co-doping strategy not only successfully increases $\Delta V/V_0$ and reduces $\Delta M$ in the deep-freezing temperature range, but also effectively weakens the energy loss caused by the hysteresis behaviors in the magneto-structural transition. In the meantime, we conducted an in-depth and comprehensive study of the physical mechanisms to regulate the $\Delta V/V_0$ and $\Delta M$ through microstructural characterizations, physical property measurements, and first-principles calculations.

## 2. Experimental and calculational methods

### 2.1. Sample synthesis

The alloy ingots of $Ni_{50}(Mn_{33-x}Cu_x)(In_{14}Ga_3)$ (x=0, 2.5, 4 and 4.5 at.%, marked as $In_{14}Ga_3$, $Ga_3Cu_{2.5}$, $Ga_3Cu_4$ and $Ga_3Cu_{4.5}$, hereafter) were prepared by arc-melting technique. The undoped $Ni_{50}Mn_{33}In_{17}$ (denoted as $In_{17}$), $Ni_{50}Mn_{34.8}In_{15.2}$ ($In_{15.2}$), and $Ni_{50}Mn_{34.5}In_{15.5}$ ($In_{15.5}$) alloys were also synthesized as comparative reference samples. The ingots were melted four times for composition homogenization. Considering the high volatility of Mn compared with other components, 1 wt.% excess Mn was added to compensate for the loss during the melting process. The directional solidification was performed using the liquid-metal-cooling technique. During the directional solidification, the arc-melted ingots were remelted and injected into a copper mold with a diameter of 10 mm under the protection of argon. Then, the suction casting bars were melted and directionally solidified in the corundum tube using the Bridgman method with a growth rate of 0.05 mm s$^{-1}$. These directionally solidified bars were sealed in a quartz tube and homogenized at 1173 K for 24 h, and then quenched into water.

*2.2. Characterization of magnetocaloric effect*

Magnetization: The temperature- and field-dependent magnetization measurements, $M(T)$ and $M(\mu_0H)$, were performed in a superconducting quantum interference device (MPMS, Quantum Design). The $M(T)$ curves upon cooling and heating were recorded at a scan rate of 5 K min$^{-1}$. The $M(\mu_0H)$ curves were measured using a discontinuous heating loop method with a maximum applied field of 5 T, *i.e.*, before measuring each $M(H)$ curve, the sample was first cooled down to 180 K under zero-field to ensure that it was in a pure martensitic state, and then heated up to the testing temperature, $T_{test}$. At each $T_{test}$, the magnetic field-up and field-down cycles were continuously applied two times to evaluate the reversible MCE.

Magnetic entropy change: The isothermal magnetocaloric entropy change $\Delta S_M$ values induced by a field of 5 T during the first and second field cycles were estimated by using the Clausius-Clapeyron relation [20, 24, 25], $\Delta S_M = -\Delta M' \times (d\mu_0 H_{cr}/dT)$, where $\Delta M'$ refers to the magnetization variation from the initial to the final field and $d\mu_0 H_{cr}/dT$ stands for the temperature sensitivity of the critical magnetic field, $\mu_0 H_{cr}$, that induces the transformation. The selection of the Clausius-Clapeyron method is to avoid the possible numerical error in estimating the reversible $\Delta S_M$ in other approaches due to the co-existence of martensite and austenite during the reverse transformation. During the calculations, the magnetization at $\mu_0 H$ of 0.4 T was chosen as the initial value to avoid the error caused by numerical instability.

Adiabatic temperature change: Direct measurement of $\Delta T_{ad}$ was performed using both continuous heating and discontinuous heating modes in a homemade experimental setup with a maximum field of 1.5 T produced by NdFeB permanent magnet. The temperature variation during the magnetic field-induced inverse martensitic transformation was monitored by a thermocouple attached to the sample surface. For the continuous heating mode, the sample was first cooled to the pure martensite state (228 K) under zero field and then continuously heated to each $T_{test}$, at which $\mu_0 H$ was applied, and the temperature signal was recorded. For the discontinuous heating mode, the sample was cooled down to 228 K in zero field before measuring each temperature point and then heated to $T_{test}$, after which the temperature signal was recorded by applying a field in situ.

Refrigeration capacity: The refrigeration capacity $RC$ was evaluated by integrating the $\Delta S_M(T)$ curve over the full width at half maximum $\Delta T_{FWHM}$, $RC=\int_{T_{cold}}^{T_{hot}} \Delta S_M dT$, where $T_{hot}$ and $T_{cold}$ were defined for $\Delta T_{FWHM}$ of the $\Delta S_M$ curve. The effective refrigeration capacity, $RC_{eff}$, was then calculated by subtracting the average hysteresis loss, $HL_{av}$, from $RC$. Here, the hysteresis loss, $HL$, was estimated by integrating the overlapped area between the field-up and the field-down $M(\mu_0 H)$ curve.

## 2.3. Characterization of elastocaloric effect

Strain-stress measurements: The superelasticity and elastocaloric properties were tested in a universal testing machine (DNS-10) equipped with a temperature control chamber. The isothermal stress-strain curves were measured at temperatures ranging from 218 K to 258 K and a strain rate of $1.0 \times 10^{-3}$ s$^{-1}$. Rectangular samples with the size of $3 \times 5 \times 8$ mm$^3$ were used in mechanical testing, where the longest axis is parallel to the solidification direction (SD) of the sample.

Elastocaloric entropy change: Based on the stress-strain curves under various temperatures, the value of isothermal elastocaloric entropy changes, $\Delta S_E$, was calculated by using the relation of $\Delta S_E = v_0 \int_0^\varepsilon \left(\frac{\partial \sigma}{\partial T}\right)_\varepsilon d\varepsilon$, where $v_0$ represents the specific volume, e.g., $1.25 \times 10^{-4}$ m$^3$ kg$^{-1}$ for the Ga$_3$Cu$_{2.5}$ compound.

Adiabatic temperature change: The value of $\Delta T_{ad}$ caused by the reverse martensitic transformation was directly measured at different temperatures with an unloading rate of $\dot{\varepsilon}_{unload}$

~0.31 s$^{-1}$ and ~0.83 s$^{-1}$. Temperature variation was recorded by a thermocouple welded on the center of the sample surface.

Coefficient of performance: The elastocaloric coefficient of performance, *COP*, which describes refrigeration efficiency, is determined by the ratio of the cooling power, $\Delta Q$, to the input mechanical work, $\Delta W$ [49, 50]. For the unloading process, $\Delta Q$ is the absorbed heat that can be estimated by the latent heat of $\Delta T_{ad} \times c_p$, and $\Delta W$ can be calculated by integrating the area of the stress hysteresis loops, $\Delta W = \oint \sigma(\varepsilon) d\varepsilon/\rho$, where $c_p$ is the heat capacity, and $\rho$ is the material density. For $Ga_3Cu_{2.5}$, $\rho$ is determined as $8.0 \times 10^3$ kg m$^{-3}$.

*2.4. Characterization of crystal structures and phase transitions*

Differential scanning calorimetry: The behaviors of magneto-structural transition were detected by differential scanning calorimetry (DSC, TA-Q100) with a sweeping rate of 10 K min$^{-1}$. The characteristic temperatures were determined from the DSC curves using the tangent extrapolation method. The specific heat capacity, $c_p$, near the martensitic transformation was measured by a modulated DSC method, which was calibrated using a standard sample of sapphire.

Neutron powder diffraction: The neutron powder diffraction (NPD) measurements were carried out at 150 K and 320 K, on the High-Intensity Powder Diffractometer, Wombat, at the Australian Organization of Nuclear Science and Technology (ANSTO). The neutron wavelength λ of 1.54 Å (for $In_{15.5}$) and 2.42 Å (for $Ga_3Cu_{2.5}$) were used. The NPD measurements were also performed at 140 K and 300 K to further confirm the occurrence of the phase transition, on the Peking University High-Intensity Powder Diffractometer (PKU-HIPD) at China Institute of Atomic Energy (CIAE) (λ=1.48 Å).

*In-situ* X-ray diffraction: Crystal structures at different temperatures were examined using the X-ray diffractometer (XRD, Rigaku SmartLab SE) with Cu-$K_\alpha$ radiation. Variable temperature measurements were carried over the temperature range from 100 K to 360 K in 10 K steps. Each spectrum is measured in the 2$\theta$ range of 20° to 120°. Refinements of X-ray diffraction were performed using the Le-Bail method with the Jana software [51]. The pole figure of {004}$_A$ was measured by the Schulz back-reflection method using the Cu-$K_\alpha$ radiation ($\lambda$=1.54059 Å) in a Rigaku SmartLab X-ray diffractometer.

Microstructure characterization: Microstructure and crystallographic preferred orientation were examined by using a scanning electron microscope (SEM, Jeol JSM 7001F) with an electron backscatter diffraction (EBSD) acquisition camera and Channel 5 software.

## 2.5. First-principles calculations

First-principles calculations were performed to clarify the origin of the Ga-, Cu-, and Ga & Cu substitution on ferromagnetism and volume change by using density functional theory (DFT) as implemented in the Vienna Ab initio Simulation Package (VASP) in combination with the projector augmented wave (PAW) method. The exchange-correlation function was described by the Perdew-Burke-Ernzerhof (PBE) parametrization of generalized gradient approximation (GGA). A plane-wave kinetic energy cutoff was set to be 600 eV. Electronic energy self-consistency was enforced up to $10^{-6}$ eV, and all-force-norm self-consistency in geometry optimization was relaxed to be $10^{-4}$ eV/Å. The valence electron configurations of Ni ($3d^94s^1$), Mn ($3d^64s^1$), Cu ($3d^{10}4s^1$), In ($5s^25p^1$) and Ga ($4s^24p^1$) were adopted. In addition, to realize a high calculation efficiency, this work adopted two approximations for crystal structure of martensite and transformation strain path: 1) the tetragonal structural model (space group: *I*4/*mmm*) was utilized to describe the lattice of martensite; 2) The widely used Bain strain model was used to represent the strain path of the martensitic transformation [43].

The finite-temperature structural models of $Ga_3Cu_{2.5}$ and $In_{15.5}$ compounds were generated using first-principles molecular dynamics (FPMD) simulations. A 128-atom supercell with 2×2×2 $L2_1$-type unit cells was adopted for $Ga_3Cu_{2.5}$ and $In_{15.5}$. During the calculations, the canonical ensemble (NVT) was employed, and the Nosé–Hoover thermostat was applied to control the system temperature. In FPMD simulations, the equation of motion is solved through the velocity algorithm with a time step of 2 fs. At 400 K, a series of 10-ps AIMD simulations were carried out, with each simulation commencing from an isotropically expanded and contracted ground state structure.

## 3. Results and discussion

### 3.1. Transformation entropy changes and thermal hysteresis

The phase transitions were firstly tracked with DSC and magnetization measurements as shown in Fig. 1(a) and Fig. S1 in the Supplementary Materials. Then, the total phase transition entropy changes, $\Delta S_{tr}$, were estimated by $\Delta H/T_M$, where $\Delta H$ represents the enthalpy change determined by integrating the area of the exothermic peak in the DSC curves and $T_M$ is the equilibrium temperature of forward martensitic transformation defined as $(M_s+M_f)/2$. The thermal hysteresis, $\Delta T_{hys}$, was determined by the formula of $T_0^A - T_0^M$, where $T_0^M$ and $T_0^A$ are the equilibrium temperatures of the forward and reverse martensitic transformation, respectively (Fig. 1(b) and Fig. S1(b)). A comprehensive comparison of $\Delta S_{tr}$ and $\Delta T_{hys}$ is summarized in Fig. 1(c) and Fig. 1(d) [3-43]. Accordingly, the $Ga_3Cu_{2.5}$ compound exhibits a deep-freezing transformation temperature of $T_M \sim 225.0$ K, and more importantly, its $\Delta S_{tr}$ is as high as 22.0 J $kg^{-1}$ $K^{-1}$, ranking among the top in the deep-freezing temperature range (Fig. 1(c)). Meanwhile, it also has a smaller $\Delta T_{hys}$ of 3 K (Fig.

1(d)), which enables it to achieve significant reversible MCE and eCE to serve the biomedical engineering, exploration of space, and extreme environment monitoring in the deep-freezing temperature range.

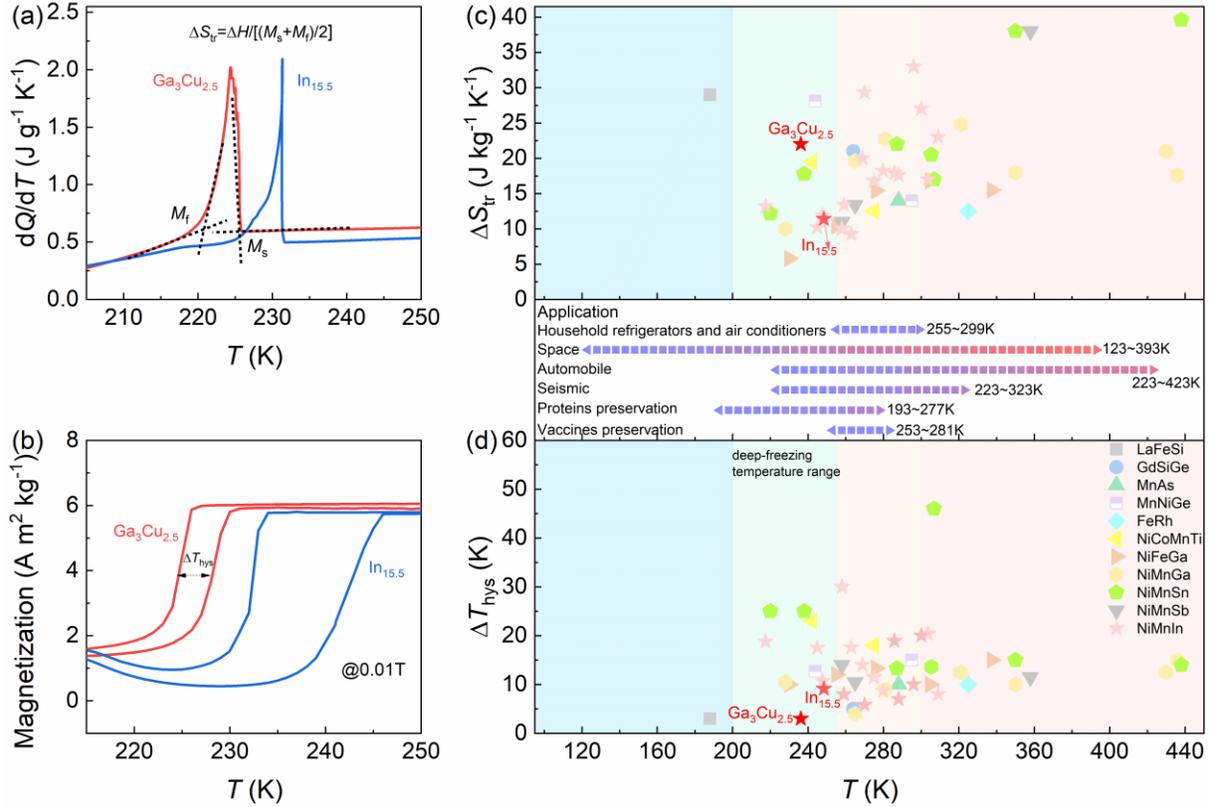

Fig. 1. **Summarization of transformation entropy changes and thermal hysteresis for the typical SLC materials.** (a) DSC curves and (b) temperature-dependent magnetization $M(T)$ curves under the magnetic field of 0.01 T of $Ni_{50}(Mn_{30.5}Cu_{2.5})(In_{14}Ga_3)$ and $Ni_{50}Mn_{34.5}In_{15.5}$ alloys. (c) Comparison of the total transformation entropy change, $\Delta S_{tr}$, and (d) thermal hysteresis, $\Delta T_{hys}$ over a broad temperature range [3-43]. $T$ represents the finish temperature of the reverse martensitic transformation $A_f$. The temperature ranges required for some special application scenarios are listed between (c) and (d) [14, 44], and this work aims to develop MCE and eCE materials in the deep-freezing temperature range from ~200 K to ~255 K to serve the biomedical engineering, exploration of space, extreme environment monitoring, etc.
- 9 -

*3.2. Magnetocaloric and elastocaloric performance*

The large $\Delta S_{tr}$ and small $\Delta T_{hys}$ predict the $Ga_3Cu_{2.5}$ compound has outstanding MCE and eCE performances. Therefore, the magnetocaloric properties, including the isothermal magnetic entropy changes, $\Delta S_M$, and the adiabatic temperature change, $\Delta T_{ad}$, were first assessed from the magnetization/demagnetization measurements (Methods and Supplementary Section 2). As summarized in Fig. S2, the maximum magnetic entropy change, $\Delta S_M^{max}$, is determined as 20.0 J kg$^{-1}$ K$^{-1}$, making it one of the best MCE materials within the deep-freezing temperature range (Table S1). In addition, as shown in Fig. 2(a), the *HL* of the $Ga_3Cu_{2.5}$ alloy is much smaller than that of the $In_{15.5}$ alloy. The effective refrigeration capacity, $RC_{eff}$, was estimated by subtracting the average *HL* ($HL_{av}$) from *RC* (Methods). The first cycle (the maximum refrigeration capacities) and second cycle (the reversible refrigeration capacities) values under $\mu_0 \Delta H$ of 5 T for $Ga_3Cu_{2.5}$ were determined as 182.1 J kg$^{-1}$ and 161.8 J kg$^{-1}$, respectively. A cross-comparison with other typical SLC-based MCE materials is displayed in Fig. 2(c) [5, 42, 52-68]. It is easy to read that the $RC_{eff}$ value for $Ga_3Cu_{2.5}$ is larger than that for most other compounds in the deep-freezing temperature range due to its relatively larger $\Delta S_M$ and smaller $HL_{av}$ (Table S1). In addition, the MCE |$\Delta T_{ad}$| of $Ga_3Cu_{2.5}$ (4 K) under low field (1.5 T) also exceeds that of most SLC-based MCE materials, as shown in Fig. 2(e) [3, 24, 25, 29, 30, 65, 66, 69-78].

Then, the elastocaloric properties of $Ga_3Cu_{2.5}$ were studied by measuring the superelasticity properties. In comparison with other typical SLC materials, the $Ga_3Cu_{2.5}$ compound exhibits a lower critical stress, $\sigma_M^s$, of 19–145 MPa, a narrower stress hysteresis, $\Delta\sigma_{hys}$, of ~17 MPa, and a much smaller hysteresis energy loss, $E_{loss}$, of ~0.93 MJ m$^{-3}$ over 223 K to 258 K (Fig. 2(b) and Fig. S4). As shown in Fig. 2(b), the $E_{loss}$ values of the $Ga_3Cu_{2.5}$ alloy are much smaller than that of the $In_{15.5}$ alloy. In addition, the experimental $\Delta T_{ad}$ cross the reverse martensitic transformation was recorded as −6.1 K to −7.0 K in a deep-freezing temperature range of 228 K to 268 K, being close to the calculated adiabatic temperature changes $\Delta T_{cal}$ of −7.4 K at 248 K, as shown in Fig. S5 (more details are given in Supplementary Section 3). More importantly, the small $E_{loss}$ of $Ga_3Cu_{2.5}$ brings about a large coefficient of performance, *COP*, for the elastocaloric effect. As summarized in Fig. 2(d), the estimated *COP* values for $Ga_3Cu_{2.5}$ are 29.0, 30.4, 33.8, and 35.0 at 228 K, 233 K, 243 K, and 253 K, respectively, ranking as the highest one among the typical elastocaloric materials [14-16, 48, 79-92]. In addition, the eCE $|\Delta T_{ad}|$ of $Ga_3Cu_{2.5}$ also exceeds that of most elastocaloric materials in the deep-freezing temperature range, as shown in Fig. 2(f) [14, 15, 80, 82, 86, 90, 92-96].

To illustrate the importance of the Ga and Cu co-doping effects in improving the magneto- and elasto-caloric effects of the NiMnIn-based compounds, the MCE and eCE properties of the undoped $In_{15.5}$ sample were also evaluated. As shown in Fig. 2, the $RC_{eff}$, *COP*, MCE $|\Delta T_{ad}|$ and eCE $|\Delta T_{ad}|$ of $In_{15.5}$ are much smaller than those of $Ga_3Cu_{2.5}$ because of its relatively smaller $\Delta S_{tr}$ but larger $\Delta T_{hys}$ of $In_{15.5}$.

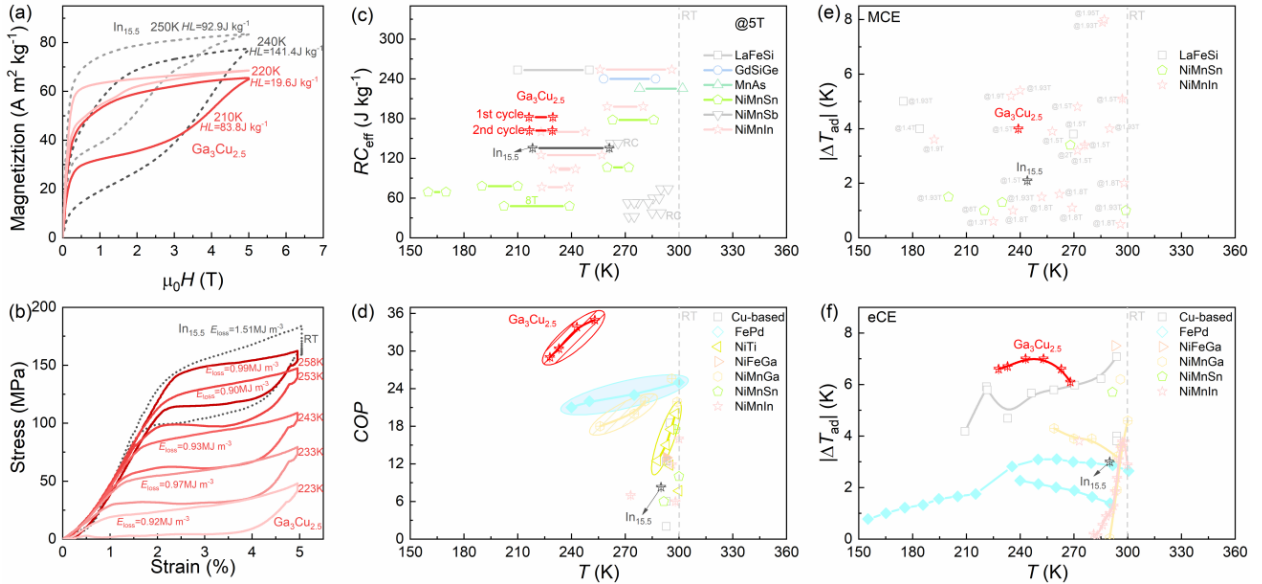

Fig. 2. **Performances of the magneto- and elasto-caloric effects of $Ga_3Cu_{2.5}$ and $In_{15.5}$.** (a) Comparison of the isothermal magnetization $M(\mu_0H)$ curves and (b) the isothermal stress-strain

curves for $Ga_3Cu_{2.5}$ alloy and $In_{15.5}$ alloy. (c) Comparison of the effective MCE refrigeration capacity $RC_{eff}$ [5, 42, 52-68] and (d) the eCE coefficient of performance, $COP$ [14-16, 48, 79-92], for $Ga_3Cu_{2.5}$ alloy and other reported caloric materials. (e) Comparison of the MCE adiabatic temperature change [3, 24, 25, 29, 30, 65, 66, 69-78] and (f) the eCE adiabatic temperature change [14, 15, 80, 82, 86, 90, 92-96] for $Ga_3Cu_{2.5}$ alloy and other reported caloric materials. The length of the bar where the $RC_{eff}$ value is located represents $\Delta T_{FWHM}$, in other words, the working temperature window. Open square, squared plus, and solid square symbols represent the non-textured polycrystalline, the textured polycrystalline, and the single crystal, respectively. Except for those marked in (c), the $RC_{eff}$ values are obtained under the magnetic field of 5 T. The magnetic field required for adiabatic temperature change in (e) is marked. Generally, the elastocaloric effects in (d) and (f) occur during complete phase transition, and the required stress or strain is not specifically noted.

*3.3. Crystal structures and microstructures*

As demonstrated above, the excellent MCE and eCE of the directionally solidified $Ga_3Cu_{2.5}$ compound mostly benefit from its large total entropy change $\Delta S_{tr}$ and small hysteresis $\Delta T_{hys}/\Delta \sigma_{hys}$ of the martensitic transformation. To expose the mechanisms behind these two key and excellent properties, we conducted a comprehensive study of the magneto-structural transitions in $Ga_3Cu_{2.5}$ and the $In_{15.5}$ reference sample through *in-situ* characterization of crystal structure and microstructures, first-principles calculations, and magnetization measurements. Following careful analyses on $\Delta V/V_0$ and $\Delta M$, it shows that the co-doping of Ga and Cu can increase the lattice contribution of $|\Delta S_{lat}|$ and reduce the magnetic contribution of $|\Delta S_{mag}|$ thereby leading to a larger $\Delta S_{tr}$.

The crystal structures of $Ga_3Cu_{2.5}$ below and above $T_M$ were first checked with NPD measurements on the WOMBAT diffractometer at the ANSTO, Australia. The NPD patterns at 150 K and 320 K are shown in Fig. 3(a), which could be indexed as a six-layered modulated monoclinic structure with a space group of $I2/m(\alpha0\gamma)00$ (marked as 6M) and a cubic structure with $Fm\bar{3}m$ (marked as $L2_1$), respectively. One obvious feature to distinguish these two symmetries in the NPD patterns is the emergence of a series of satellite peaks near the main Bragg peaks, and this phenomenon is also captured with the NPD measurements on the PKU-HIPD diffractometer at CARR, China (Fig. S7). Next, the evolution of the crystal structures with temperatures for $Ga_3Cu_{2.5}$ was tracked in detail by an *in-situ* XRD measurement (see Supplementary Section 4 and Fig. S8 for the analysis of $In_{15.5}$). The contour image of the temperature-variable XRD patterns for $Ga_3Cu_{2.5}$ is shown in Fig. 3(b). A clear martensitic transformation can be easily spotted at ~210 K, being close to the values from DSC or magnetization measurements (Fig. S1). All the XRD patterns were analyzed using Rietveld or Le-Bail refinements with the Jana2006 program [51]. Taking the patterns at 215 K and 125 K as examples, all the Bragg peaks can be well fitted with the $L2_1$ and 6M structures, as shown in Fig. 3(d) and Fig. 3(e), respectively. Finally, the lattice parameters as a function of temperature were successfully extracted from the refinement results of the XRD patterns, which will be further discussed in the next section.

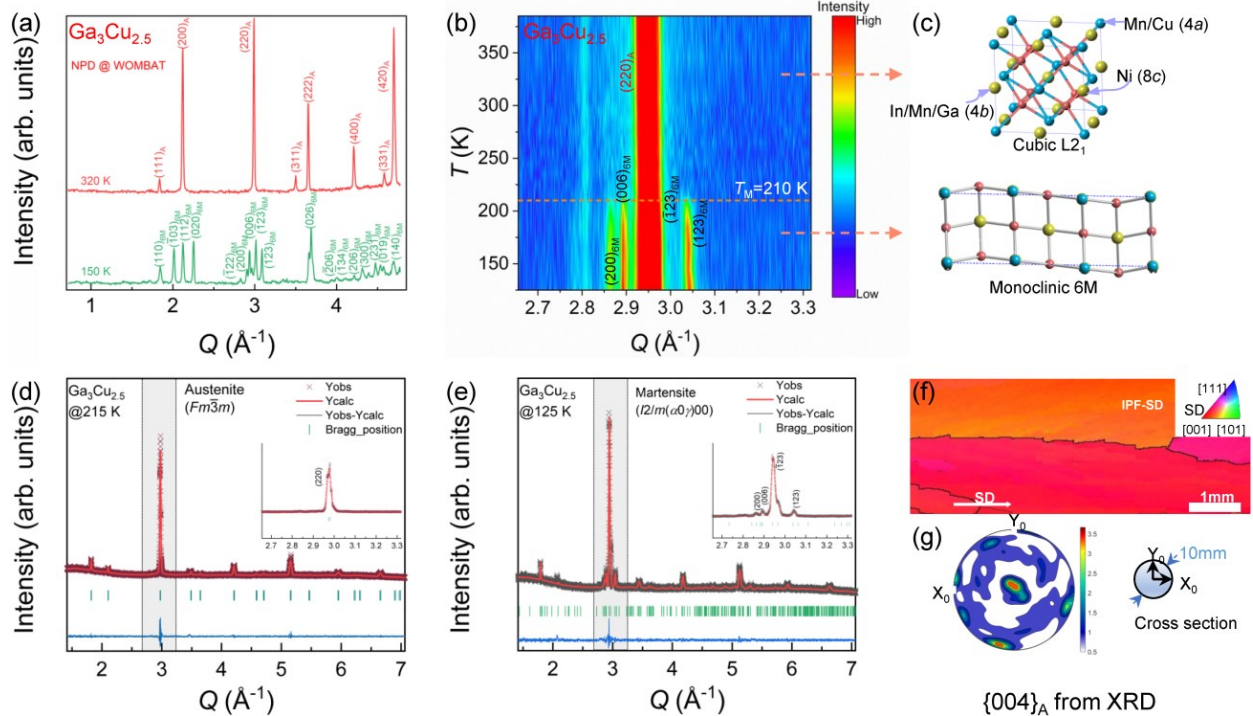

Fig. 3. **Crystal structures and microstructures of the Ga$_3$Cu$_{2.5}$ compound.** (a) The NPD patterns at 150 K and 320 K collected on the WOMBAT neutron diffractometer at ANSTO, Australia. (b) Contour images of the *in-situ* powder XRD patterns from 125 K to 385 K. (c) Crystal structures of the cubic L2$_1$ austenite with a space group of $Fm\bar{3}m$ and the monoclinic 6M martensite with $I2/m(\alpha 0\gamma)00$. The refinement results of the XRD patterns (d) at 215 K in the austenitic phase and (e) at 125 K in the martensitic phase. The vertical bars mark the Bragg peak positions of the austenitic and martensitic phases. The bottom black solid lines (Yobs−Ycalc) are the difference between observed (Yobs, black circles) and calculated (Ycal, red solid line) data. The inset in (d) and (e) display the refinement results for $Q$ ranging from 2.7 Å$^{-1}$ to 3.3 Å$^{-1}$. (f) EBSD orientation micrograph of the longitudinal section and (g) {004}$_A$ pole figure of austenite measured by XRD technique.

Besides the crystal structures, the microstructures, including the crystallographic preferred orientation and the grain size, are analyzed through EBSD and XRD macro-texture measurements. Following Fig. 3(f) and Fig. 3(g), the Ga$_3$Cu$_{2.5}$ sample has a strong preferred orientation of <001>$_A$ // SD, corresponding to the direction in which it grows most easily along the temperature gradient during directional solidification. Furthermore, the sample possesses columnar-shape grains along SD, with a long axis of 2–5 mm and a wide axis of 0.5–2 mm. Generally, this unique microstructure with coarse columnar grains and strong texture is beneficial for both eCE and MCE [30].

### 3.4. Lattice anharmonicities dominating $\Delta V/V_0$ and $|\Delta S_{lat}|$

During an SLC phase transition, the lattice entropy change $|\Delta S_{lat}|$ makes a main contribution to $\Delta S_{tr}$ and its magnitude correlates to the unit cell volume change ratio, $\Delta V/V_0$ [97]. Therefore, based on the above refinement results of the XRD patterns, we analyzed the lattice parameters and the unit cell volume for Ga$_3$Cu$_{2.5}$ and In$_{15.5}$ alloys in Fig. 4(a)–4(d). The value of $\Delta V/V_0$ for Ga$_3$Cu$_{2.5}$ is determined as ∼1.39%, which is much larger than the value of ∼0.85% for In$_{15.5}$ and also ranked as one of the largest in the Ni–Mn-based alloys (Fig. 4(c) and 4(d)).

This large value of $\Delta V/V_0$ in Ga$_3$Cu$_{2.5}$ during the magnetic martensitic transformation raises two fundamental questions. Firstly, is it Ga or Cu doping that determines the magnitude of $\Delta V/V_0$? To answer this question, we compared the effects of Ga doping, Cu doping, and Ga-Cu co-doping on the unit cell volumes of $V_M$ and $V_A$ in the martensitic and austenitic phases by first-principles calculations. As displayed in Fig. 4(e), all of Ga, Cu, and Ga-Cu doping can reduce $V_M$ and $V_A$. However, $V_M$ decreases faster than $V_A$ with increasing doping content. Therefore, all three doping methods can enlarge $\Delta V/V_0$. Nonetheless, we find that the discrepancy in the doping concentration sensitivity of $V_M$ and $V_A$ is larger for Cu doping than for Ga doping, *e.g.*, (d$V_M$/dy)/(d$V_A$/dy)=∼3.03 for Cu doping while (d$V_M$/dx)/(d$V_A$/dx)=∼1.05 for Ga doping. Therefore, it is the Cu dopant that makes the main contribution to the larger $\Delta V/V_0$ value in Ga$_3$Cu$_{2.5}$ (Fig. 4(f) and Fig. S9). In other words, Cu doping is a more effective way to amplify $\Delta V/V_0$.

The second question is what happens to the unit cell after doping? To unveil the details, the temperature dependences of $V_M$ and $V_A$ are investigated in Fig. 4(c). $V_M$ is smaller than $V_A$ in both Ga$_3$Cu$_{2.5}$ and In$_{15.5}$ compounds, and both $V_M$ and $V_A$ expand with increasing temperature. In addition, $V_A$ of the two samples has similar thermal expansion coefficients of $3.44\times10^{-5}$ K$^{-1}$ (Ga$_3$Cu$_{2.5}$) and $3.43\times10^{-5}$ K$^{-1}$ (In$_{15.5}$), respectively. However, $V_M$ of Ga$_3$Cu$_{2.5}$ ($1.40\times10^{-4}$ K$^{-1}$) expands slower than that of In$_{15.5}$ ($2.33\times10^{-4}$ K$^{-1}$), making $V_M$ in Ga$_3$Cu$_{2.5}$ not as easy to approach $V_A$ as in In$_{15.5}$, which then leads to larger $\Delta V/V_0$ and $|\Delta S_{lat}|$ in Ga$_3$Cu$_{2.5}$.

According to solid-state theory, thermal expansion arises from anharmonic lattice vibrations. The lattice anharmonicity can be quantified using Grüneisen parameter [98, 99]:

$$\gamma_i = -\frac{V}{\omega}\frac{\partial \omega}{\partial V} \qquad (1)$$

where $\omega$ is the phonon frequency. To demonstrate the different lattice anharmonicity of the martensitic phase in the Ga$_3$Cu$_{2.5}$ and In$_{15.5}$ compounds, the thermal expansion coefficients and Grüneisen parameters are simulated by first-principles calculations in Fig. 4(g). Although the absolute values are different from the experimental values (which is attributed to the description of volume-dependent thermal effects within the quasi-harmonic approximation in first-principles calculations to simulate anharmonic influences), the trend of $\gamma$ for In$_{15.5}$ is larger than that for Ga$_3$Cu$_{2.5}$ is successfully captured. To facilitate understanding of lattice anharmonicity, we further analyzed and compared the phonon (lattice vibration) potential wells of the two samples. As shown in Fig. 4(h) and Fig. 4(i), the phonon potential wells of both samples deviate from the parabolic (quadratic or harmonic) behavior. However, the phonon potential well of the In$_{15.5}$ compound deviates more significantly, and its fitting needs to consider the quartic term, showing a larger anharmonicity, which is consistent with the analysis result of the Grüeneisen constant. In short, the larger $\Delta V/V_0$ and $|\Delta S_{lat}|$ in Ga$_3$Cu$_{2.5}$ essentially stem from the weaker lattice anharmonicity of the low-$T$ martensitic phase.

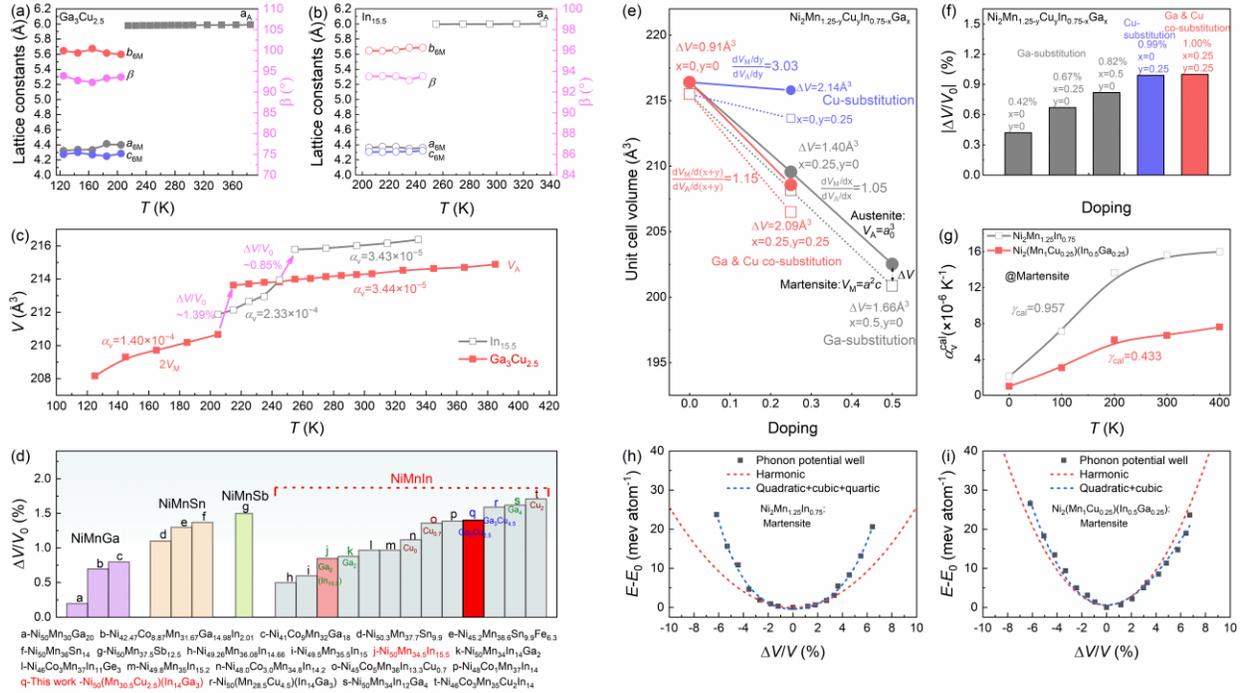

**Fig. 4. Origin of different $\Delta V/V_0$ values in the Ga$_3$Cu$_{2.5}$ and In$_{15.5}$ compounds.** Variations of the lattice parameters for (a) the Ga$_3$Cu$_{2.5}$ and (b) the In$_{15.5}$ compounds as determined from the *in-situ* XRD measurements, respectively. (c) Evolutions of the unit cell volumes for Ga$_3$Cu$_{2.5}$ and In$_{15.5}$. The volume thermal expansion coefficients, $\alpha$, of the martensitic and austenitic phases and the volumetric expansion ratios, $\Delta V/V_0$, across the martensitic transformation are marked. (d) Comparison of $\Delta V/V_0$ for Ga$_3$Cu$_{2.5}$ with some typical SLC Ni−Mn-based alloys: a[32], b[33], c[34], d,e[35], f[36], g[37], h[38], i[39], j: this work (In$_{15.5}$), k[28], l[24], m[40], n[26], o[41], p[42], q: this work (Ga$_3$Cu$_{2.5}$), r[43], s[28] and t[25]. Evolutions of (e) the unit cell volume for the austenite and martensite and (f) the volume change during transition of Ni$_2$(Mn$_{1.25-y}$Cu$_y$)(In$_{0.75-x}$Ga$_x$) alloys by ab-initio calculations. (g) The calculated volume thermal expansion of martensite in undoped ternary and Ga & Cu co-doped alloys. (h,i) Anharmonic phonon potential wells calculated from DFT for Ga$_3$Cu$_{2.5}$ and In$_{15.5}$ compounds, and the quadratic (harmonic), cubic and/or quartic fits to the phonon potentials.

*3.5. Tailoring spin polarization to relieve negative |ΔS$_{mag}$|*

This section delves into the magnetic contribution $|\Delta S_{\text{mag}}|$ and the magnetization difference $\Delta M$ over the magneto-structural transition. Fig. 5(a) and 5(b) display the temperature-dependent magnetization $M(T)$ and isothermal magnetization $M(\mu_0 H)$ curves for both the undoped ternary and Ga & Cu co-doped alloys. In the undoped samples, the magnetization of the ferromagnetic austenitic phase increases as the temperature decreases, resulting in an increase in $\Delta M$ from 83.1 A m$^2$ kg$^{-1}$ at 255.2 K for In$_{15.2}$ to 86 A m$^2$ kg$^{-1}$ at 210.0 K for In$_{15.5}$ under a $\mu_0 H$ of 5 T, impeding efforts to improve $\Delta S_{\text{tr}}$. This trend is true for most NiMnIn-based compounds [43, 45]. Nevertheless, the co-doping of Ga and Cu breaks this rule. It not only successfully lowers $T_M$ to 212.0 K (5 T) but also markedly suppresses $\Delta M$ to 51.5 A m$^2$ kg$^{-1}$. The reduction in $\Delta M$ is attributed to a slight uptick in the net magnetization of the martensite and, notably, a substantial decrease in the magnetization of the austenite, as further illustrated by the saturated magnetizations in Fig. 5(b). A comprehensive comparison of the key parameters of the magneto-structural transitions is outlined in Table S2, underscoring once more the significant role of introducing Cu and Ga in synergetically lowering $T_M$ and raising $\Delta S_{\text{tr}}$.

To elucidate the mechanism of the Ga-Cu co-doping in modulating the magnetization in austenite, we conducted first-principles calculations. The majority-spin and minority-spin density of states (DOS) and their integrated values (IDOS) for the austenitic phase are carefully examined and compared. In the case of Ga-doping, the IDOSs at the Fermi level show no significant variations with the Ga content (Fig. 5(c)). On the other hand, the introduction of Cu clearly adjusted the asymmetry between the majority and minority DOSs. Specifically, the difference between the majority and minority IDOSs at the Fermi level decreases from 3.18 states in $Ni_2Mn_{1.25}In_{0.75}$ to 2.22 states in $Ni_2(Mn_1Cu_{0.25})In_{0.75}$, and further to 2.19 states in $Ni_2(Mn_1Cu_{0.25})(In_{0.5}Ga_{0.25})$, as illustrated in Fig. 5(d). This observed trend is further supported by the comparison between $Ni_2Mn_{1.5}In_{0.5}$ and $Ni_2Mn_{1.25}Cu_{0.25}In_{0.5}$ (Fig. S10). Furthermore, a careful comparison of the DOSs indicates that the variation of the spin-polarized DOSs' asymmetry with Cu doping primarily stems from changes in the majority- and minority-spin states around −3.0 eV and −0.9 eV.

The atomically resolved magnetic moments for the Ga & Cu doped austenitic phase were further assessed through partial DOSs analysis. As illustrated in Fig. S11, the variation in the majority- and minority-spin states are mainly associated with 8$c$ (Ni) and 4$a$ (Mn) Wyckoff positions. Then, the magnetic moments for each atomic position among different $Ni_2(Mn_{1.25-y}Cu_y)(In_{0.75-x}Ga_x)$ samples are compared in Fig. 5(e) and 5(f). This comparison once again highlights significant changes in the magnetic moments at the 8$c$ (Ni) and 4$a$ (Mn) positions. The total magnetic moment, $M_{total}$, notably decreases from 3.16 $\mu_B$ f.u.$^{-1}$ in the undoped $Ni_2Mn_{1.25}In_{0.75}$ to 2.17 $\mu_B$ f.u.$^{-1}$ in Ga-Cu co-doped $Ni_2(Mn_1Cu_{0.25})(In_{0.5}Ga_{0.25})$ alloy. In short, the reductions in $\Delta M$ and $|\Delta S_{mag}|$ in the Ga-Cu do-doped samples are attributed to the modification of the spin-polarized DOSs, which mainly lie in the 4$a$ position.

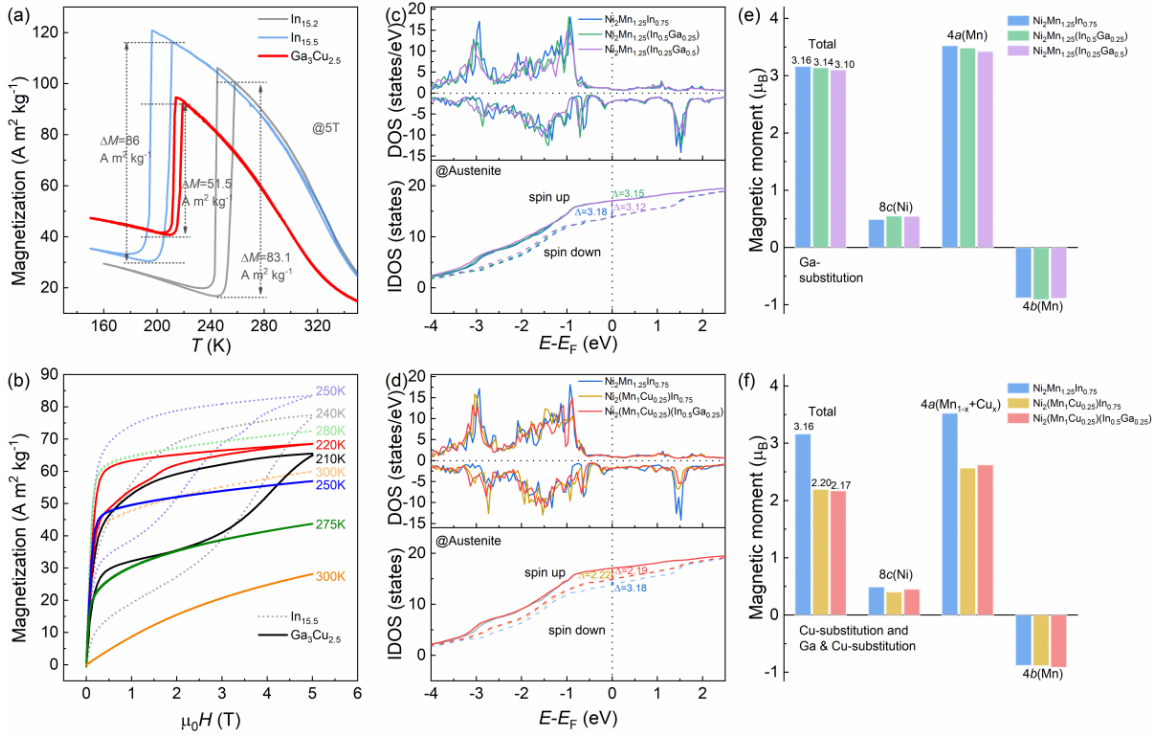

Fig. 5. **Magnetizations and modified spin-polarized DOSs in the $Ni_2(Mn_{1.25-y}Cu_y)(In_{0.75-x}Ga_x)$ compounds.** (a) Temperature-dependent magnetization $M(T)$ curves and (b) isothermal magnetization $M(\mu_0 H)$ curves for the undoped ternary and Ga & Cu co-doped alloys. Spin-polarized density of states (DOS) and the integrated density of states (IDOS) of the austenite for (c) Ga-, (d) Cu-, and Ga & Cu co-doped compounds. The positive side represents the majority-spin states and the negative side represents the minority-spin states. Comparison of total and atom-resolved magnetic moments for the austenite of (e) Ga-, (f) Cu-, and Ga & Cu co-doped compounds.

*3.6. Weaken lattice mismatch giving rise to small hysteresis*

In addition to the large $\Delta S_{tr}$, the excellent magnetocaloric $RC_{eff}$ and elastocaloric $COP$ also strongly benefit from the smaller hysteresis losses (Fig. 1(b) and Fig. 2). It is worthwhile to study the reasons behind the smaller $\Delta T_{hys}$ and/or $\Delta \sigma_{hys}$ in $Ga_3Cu_{2.5}$ alloy relative to $In_{15.5}$. Geometric compatibility between the martensitic and austenitic phases is a key factor affecting the hysteresis, and good compatibility always leads to small hysteresis. The geometric compatibility during a martensitic transformation can be evaluated from different aspects, including the middle eigenvalue, $\lambda_2$, of the martensitic transformation stretch matrix $U$ and its norms of $X_\text{I}=|U^{-1}\hat{e}|$ and $X_\text{II}=|U\hat{e}|$ (where $\hat{e}$ is the unit vector of the two-fold axes for type-I and type-II twins) [100], the dissipated energy proportional to $T_\text{C}-T_\text{M}$ [26], as well as the chemical pressure generated by lattice contraction [24]. In the picture of cofactor theory, good geometric compatibility requires $\lambda_2$, $X_\text{I}$, and $X_\text{II}$ close to 1 [100].

The $\lambda_2$, $X_\text{I}$, $X_\text{II}$, $T_\text{C}-T_\text{M}$, and $V_\text{A}$ parameters of the $Ga_3Cu_{2.5}$ and $In_{15.5}$ compounds are carefully analyzed and compared in Fig. 6. Accordingly, the $\lambda_2$, $X_\text{I}$, and $X_\text{II}$ values for $Ga_3Cu_{2.5}$ are overall closer to 1 than those for $In_{15.5}$. In addition, the $T_\text{C}-T_\text{M}$ for $Ga_3Cu_{2.5}$ reads as 60.3 K, which is much smaller than the value of 82 K for the $In_{15.5}$ alloy and also smaller than the values reported for other NiCoMnIn alloys (Fig. 6 and Table S2) [45]. The smaller atomic radii of Cu and Ga than those of Mn and In also lead to a lattice contraction, e.g. $V_\text{A}$ ~213.6 Å$^3$ in $Ga_3Cu_{2.5}$ compared to $V_\text{A}$ ~215.8 Å$^3$ in $In_{15.5}$, which relieves the lattice misfit ($In_{15.5}$: $|\varepsilon_v|=0.0047$, $Ga_3Cu_{2.5}$: $|\varepsilon_v|=0.0045$) and the elastic energy between the two phases [24]. All these parameters demonstrate that the geometric compatibility in $Ga_3Cu_{2.5}$ is better than that in $In_{15.5}$, thus resulting in smaller hysteresis.

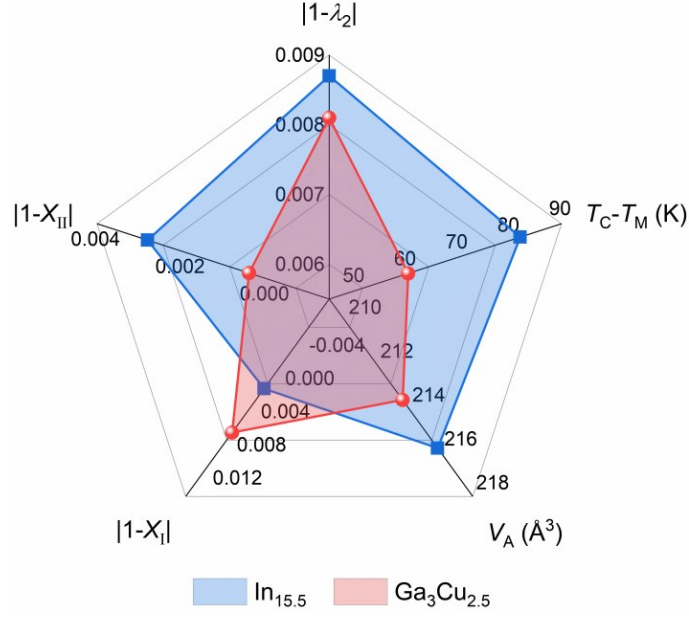

Fig. 6. **Geometric compatibility analysis of the Ga & Cu co-doped and In$_{15.5}$ compounds.** The comparison of $|1-\lambda_2|$, $|1-X_I|$, $|1-X_{II}|$, $T_C-T_M$, and unit cell volume of austenitic phase.

## 4. Conclusions

In summary, we have successfully developed a high-performance magneto- and elasto-caloric material for deep-freezing applications in the NiMnIn-based spin-lattice-coupled Heusler alloys using a multi-element alloying approach. Specifically, *via* partial substituting Ga and Cu for In and Mn, carefully balancing the In content, and creating textured microstructures, we have simultaneously modified the lattice anharmonicity, thermal expansion coefficient and electronic density of states, achieving increases in $\Delta V/V_0$ and $|\Delta S_{lat}|$, reduction in $\Delta M$ and the negative contribution from $|\Delta S_{mag}|$, as well as alleviation the hysteresis losses. These modifications in the first-order magneto-structural transitions yield excellent reversible MCE and eCE, with a huge effective magnetocaloric refrigeration capacity of ~182 J kg$^{-1}$ at a driving field of 5 T and a large adiabatic temperature change of −4 K under a low field of 1.5 T, a substantial elastocaloric coefficient of performance of ~30 and a large adiabatic temperature change of −7 K under a strain of 5% in the deep-freezing temperature range.

**Declaration of Competing Interest**

The authors declare that they have no known competing financial interests or personal relationships that could have appeared to influence the work reported in this paper.

**Acknowledgments**

This work is supported by the National Natural Science Foundation of China (Grant No. 12474024, 52101236, 52274379, 52471003, 52401252 and 52301248), Guangdong Basic and Applied Basic Research Foundation (grant no. 2021B1515140014), the Fundamental Research Funds for the Central Universities (Grant No. N2202015), the 111 Project of China (Grant No. BP0719037 and B20029), Postdoctoral Science Foundation of China (Grant No. 2024T170502 and 2024M751757), and the Guangdong Provincial Key Laboratory of Extreme Conditions. Q.R. thanks Pengfei Luo from Shanghai University for the assistance in the in-situ XRD measurements.